# Tantalum thin films sputtered on silicon and on different seed layers: material characterization and coplanar waveguide resonator performance

Moritz Singer
*School of Computation, Information and Technology*
Technical University of Munich
Garching, Germany
moritz.singer@tum.de

Benedikt Schoof
*School of Computation, Information and Technology*
Technical University of Munich
Garching, Germany

Harsh Gupta
*School of Computation, Information and Technology*
Technical University of Munich
Garching, Germany

Daniela Zahn
*Fraunhofer Institute for Electronic Microsystems and Solid State Technologies EMFT*
Munich, Germany

Johannes Weber
*Fraunhofer Institute for Electronic Microsystems and Solid State Technologies EMFT*
Munich, Germany

Marc Tornow
*School of Computation, Information and Technology*
Technical University of Munich;
*Fraunhofer Institute for Electronic Microsystems and Solid State Technologies EMFT*
Munich, Germany

*Abstract*—Superconducting qubits are a promising platform for large-scale quantum computing. Besides the Josephson junction, most parts of a superconducting qubit are made of planar, patterned superconducting thin films. In the past, most qubit architectures have relied on niobium (Nb) as the material of choice for the superconducting layer. However, there is also a variety of alternative materials with potentially less losses, which may thereby result in increased qubit performance. One such material is tantalum (Ta), for which high-performance qubit components have already been demonstrated. In this study, we report the sputter-deposition of Ta thin films directly on heated and unheated silicon (Si) substrates as well as onto different, nanometer-thin seed layers from tantalum nitride (TaN), titanium nitride (TiN) or aluminum nitride (AlN) that were deposited first. The thin films are characterized in terms of surface morphology, crystal structure, phase composition, critical temperature, residual resistance ratio (RRR) and RF-performance. We obtain thin films indicative of pure alpha-Ta for high temperature (600°C) sputtering directly on silicon and for Ta deposited on TaN or TiN seed layers. Coplanar waveguide (CPW) resonator measurements show that the Ta deposited directly on the heated silicon substrate performs best with internal quality factors $Q_i$ reaching $1\times10^6$ in the single-photon regime, measured at $T$=100 mK.

I. INTRODUCTION

Superconducting qubits are elements of circuit quantum electrodynamics (circuit QED) where a quantum two-level system can be realized. This so-called artificial atom can be used as a quantum bit or qubit for quantum computing. Today, there are several platforms known to realize qubits. Superconducting quantum bits are promising candidates as they are relatively easy scalable [1], [2]. A widely used structure, the transmon, comprises a capacitance and a Josephson junction, which provides the non-linearity that is necessary to create an anharmonic potential with energy levels that are not equally spaced. This way, transitions between specific energy levels can be addressed [3]. Another integral part of the circuitry are resonators, which can be realized as coplanar-waveguide (CPW) structures. These resonators are used to couple microwave signals into the circuit which enables initialization, control and readout of the qubits [4], and for their mutual coupling. The common materials that are mostly used for fabricating the Josephson junctions are aluminum and aluminum-oxide while for the resonators and the rest of the circuits, niobium is predominantly used. For the most part, these materials are chosen because of the large body of experience with design, fabrication and good-quality qubit operation. However, it is assumed that these materials might not be ideal in terms of qubit performance [5], [6]. In order to perform computations on a quantum computer, the qubit states need to have as high as possible coherence times. From a material perspective it is believed that, at the temperatures where the qubits are operated, two-level system (TLS) losses are the main limiting factor for the performance [7]. The origin of these TLS is not yet fully understood but it is suspected that they could originate from tunnelling atoms, dangling bonds, hydrogen defects or collective motions of small atomic groups at the various interfaces and in the bulk material [8]. Tantalum has been proposed and demonstrated as a promising alternative material for superconducting qubits [9], [10]. It is anticipated that the native oxide of tantalum is more stable and hosts less TLS than that of niobium and therefore, the performance of the qubits can be enhanced [11], [12].

Tantalum comes in two crystalline phases, the stable body-centered cubic (bcc) crystal structure phase, which is called alpha-Ta, and the meta-stable tetragonal crystal structure beta-Ta phase. The alpha-phase has a lower resistivity of 20-80 µΩ.cm while the beta-phase exhibits higher resistivity values of 150-200 µΩ.cm. So far, is has been assumed that the alpha-phase is more favourable in terms of lower TLS-losses, and the influence of various sputtering parameters on the Ta thin film properties and phase composition has been studied in this regard [13], [14], [15], [16]. There are reports that the alpha-phase can form under a certain pressure [17] and that heating above 750°C would transform beta- to alpha-tantalum [18], [19], [20]. Alpha-Ta in thin films was directly obtained

Funded through the Munich Quantum Valley project by the Free State of Bavaria, Germany.

XXX-X-XXXX-XXXX-X/XX/$XX.00 ©20XX IEEE

by depositing tantalum at elevated temperatures above 500°C [21]. High temperature sputtered Ta thin films were already used to fabricate qubits on sapphire [9], [10], [22] and superconducting CPW resonators on silicon [23], as well as to encapsulate niobium resonators [24]. An alternative strategy to obtain predominantly alpha-Ta is using a few nanometres thick seed layer before Ta deposition, thereby nucleating the formation of the alpha-Ta phase. Suitable seed layers have been demonstrated to be tantalum-nitride, niobium and titanium. Resonators built from tantalum thin films that were deposited on a niobium seed layer on silicon [25] have already been reported. By reaching Q-factors up to $2 \times 10^7$ at the large-photon-number limit, the latter work demonstrated that tantalum on seed layers is a very promising approach for fabricating high-performance resonators and qubits. As of yet, there is no known thorough comparison of the performance of alpha-Ta vs. beta-Ta to clarify whether a pure alpha-phase would provide indeed the best tantalum thin-films featuring the lowest losses. One comparison of differently deposited alpha-Ta films (high-temperature deposition vs. formation on a Nb seed layer [26]) was reported, but a systematic variation of different seed layers has not been carried out or tested for applications in superconducting qubits, to the best of our knowledge.

In this work, we report on the deposition and characterization of tantalum films using different temperatures and seed layers. Films have been deposited directly on silicon at either room temperature or at 600°C. We used tantalum-nitride, aluminum-nitride and titanium-nitride as seed layers before tantalum deposition. All deposition was done on a silicon <100> substrate. We characterized the electrical parameters of these films in terms of resistivity, residual resistance-ratio (RRR) and critical temperature $T_c$. The crystallography of the films was investigated with grazing incidence X-ray diffraction (XRD) to determine their phase composition. The films were then used to fabricate CPW resonators in the 4-8 GHz regime and to measure their internal quality factors at different power down to the single-photon regime, at a temperature of 100 mK.

## II. EXPERIMENTAL

### A. Deposition of tantalum thin films

All thin films were deposited on high-ohmic (> 10 kOhm.cm) silicon <100> substrates, which were cut from 4-inch wafers into 10 mm x 6 mm chips. Prior to deposition, the silicon chips were cleaned by ultrasonication in acetone and isopropanol for three minutes, each. After that, the substrates were dipped for 60 s into buffered oxide etch (BOE, 7:1 HF:NH4F), in order to remove the native oxide on the silicon, and then rinsed with DI-water and dried with nitrogen [*Caution: hydrofluoric acid (HF(aq)) is very hazardous to health; special care/training is mandatory*]. After the chemical cleaning steps, the substrates were mounted into the chamber of an RF-magnetron sputtering tool (Alcatel, A450). Tantalum was then deposited with a target-to-chip distance of ca. 7 cm, at 20 sccm Ar-flow, 18 µbar gas pressure and a power of 200 W. For the tantalum thin films on top of seed layers, a titanium and an aluminum target were previously added to the sputtering chamber. A 2 sccm flow of nitrogen was added during the sputtering process to obtain tantalum nitride, titanium nitride and aluminum nitride seed layers. After the deposition of a certain seed layer, the chips stayed in the vacuum of the sputtering chamber, and they were rotated towards the tantalum target such that the subsequent tantalum deposition could be carried out.

Prior to the fabrication of the actual thin-films for our material and process studies, the thicknesses and sputter rates for each of the materials were determined by sputtering on silicon chips comprising a lithographic photoresist pattern, followed by a lift-off in acetone. The step heights of the films were then measured with either a profilometer or an atomic force microscope (AFM), and the sputter time was optimized to obtain Ta films with a thickness of about 200 nm. The seed layers were optimized so that they had a thickness of 5-6 nm. Their thickness was characterized by using the same method. The height of the seed layer is most likely not precisely the same for the films where tantalum is sputtered on top. This is because, for the tantalum thin films the vacuum is not broken and tantalum is deposited immediately after seed layer deposition. For the seed layer height measurements, only the seed layer is deposited and taken out of the vacuum to do the measurement where however, some oxidation in ambient air would follow, which may likely result in the formation of 1-2 nm thick native oxides on the surface. However, the measurements still give a good upper limit estimate of the sputter rates for each material.

### B. Characterization methods

The surface of the planar tantalum thin films was investigated with an AFM. In tapping mode, pictures of area 5x5 µm² and 1x1 µm² were taken with a scan rate of 1 Hz and 512 samples per line. Subsequently, the root mean square roughness of the films was extracted from the AFM pictures. For crystallographic characterization, the planar chips were measured by X-ray diffraction (XRD). Initial measurements were done with a standard powder XRD device. However, with this method the obtained XRD spectra were strongly superimposed by silicon <100> background peaks. Because of that, grazing incidence X-ray diffraction (GI-XRD) was used instead. In GI-XRD, the angle of the X-ray source with respect to the sample surface is small and kept constant to minimize the background signal and maximize the signal from diffraction on the surface. It was determined that a fixed angle of 0.5° gives the best results for the tantalum thin films and all spectra were taken with this fixed angle of the source while the detector moved over the angular scanning range.

For electrical characterization, one chip of each kind was structured into a Hall bar. For this purpose, a design was created with a bar of 300 µm width and 3600 µm length. At defined positions that are 1500 µm apart, pads are connected to each side of the bar as voltage probes. The corresponding design was then patterned onto the chip by photolithography and a reactive ion etching (RIE) process (Oxford PlasmaPro 80 Cobra, CF4 gas, HF power 70 W, ICP power 150 W, 2 minutes etch time). Four-point resistance measurements were carried out by either placing a probe needle or a bond wire on each side of the Hall bar and on each voltage pad. A current of 100 µA was forced through the bar and the voltage drop was measured between the pads 1500 µm apart. The resistance $R$ is calculated by using Ohm's law and the resistivity $\rho$ by using

$$R = \rho \frac{l}{A} = \rho \frac{l}{w*t} \quad (1)$$

with the length $l$, the width $w$ and the thickness $t$ of the bar. The product of the latter two is the cross-sectional area $A$ of the bar, perpendicular to the current direction. The same structure was used to carry out critical temperature ($T_c$) measurements. For these measurements, the chips were bonded and mounted into an adiabatic demagnetization cryostat (kiutra, Germany) and the resistance of the thin films was measured while sweeping the temperature from 300 mK to 8 K and back to 300 mK. The temperature was swept up and down to see if the superconducting transition shows any difference depending on direction, which would indicate that the sample has a delay in thermalizing. Additionally, the resistance of the structures was measured in a four-needle probe station at room temperature. From the resistance values at room temperature and closely above the superconducting transition, the residual resistance ratio (RRR) was calculated.

In order to characterize the sputtered films in terms of performance for superconducting qubits, coplanar waveguide (CPW) resonators were fabricated. A "hanger" design with nine resonators in the frequency range 4 to 8 GHz, capacitively coupled to a feed line, was chosen and patterned by photolithography and RIE etching. The recipe was optimized such that it etches 10 nm into the silicon to guarantee that the superconducting film is etched through. After the etching, the resist is stripped off the chip. To remove the native oxide from the tantalum and the now exposed silicon in the resonator trenches, the chip is dipped in BOE for 10 minutes. After that, the chip is bonded and encapsulated into a custom-made copper box, mounted into the cryostat and cooled down to 100 mK. At both ends of the copper box SMA connectors are mounted which are used to connect to the wiring through the cryostat. A vector network analyzer P5002B (Keysight) is used for S21-measurements by sending RF-signals into the chip and detecting the transmitted signal. From the obtained spectra, the resonance peaks are selected and a fine peak sweep around these frequencies is carried out. After this, each resonance frequency is power-swept with a power ranging from -90 dBm to -160 dBm in -10 dBm steps. The results are fitted with a circle fit to extract the internal and external quality factors $Q_i$ and $Q_{ext}$, respectively, from the total quality factor $Q_{total}$ of the resonators

$$\frac{1}{Q_{total}} = \frac{1}{Q_i} + \frac{1}{Q_{ext}} \qquad (2)$$

in order to characterize the thin films with regard to their suitability for application in superconducting qubit circuits. By design, our resonators operate in the over-coupled regime, i.e., $Q_i \gg Q_{ext}$.

### III. RESULTS AND DISCUSSION

#### A. Surface morphology and crystallography

A 1 μm by 1 μm AFM picture of each investigated thin film can be seen in Figure 1. The tantalum sputtered on the aluminum nitride seed layer (a) has a RMS roughness of about 0.6 nm and appears to have a similar surface morphology as the one deposited directly on silicon at room temperature 20°C (b) with a roughness of ~0.7 nm, while the thin film sputtered on the 600°C heated substrate (c) shows an increased roughness of ~1.8 nm and elongated structures of size 100-200 nm. Tantalum sputtered on titanium nitride (d) and tantalum nitride (e) shows a RMS roughness of about 1.2 nm and 1.1 nm, respectively. The surface structure of the Ta 20°C and the Ta on AlN appears to be similar and the surface of the Ta on TiN and TaN looks to be of similar kind as well.

The GI-XRD spectra for all films were taken for a range of 20-90° of 2 theta. At the incidence beam angle of 0.5°

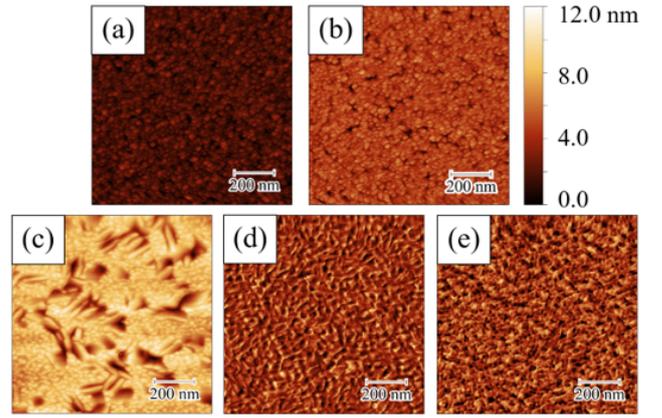

*Figure 1: AFM pictures of deposited Ta thin films sputtered on (a) an AlN seed layer, (b) the silicon substrate directly (c) a 600°C heated silicon substrate (d) a TiN seed layer and (e) a TaN seed layer. All Ta layers except the one shown in c) were sputtered at room temperature.*

almost no background signal is present. Only a very small ripple at around 52° corresponds to the silicon substrate but is barely visible, as intended by the optimization of the incidence angle setting. In Figure 2 the spectra for the investigated tantalum thin films are shown. The curve on top (a) in magenta is for the Ta thin film on the AlN seed layer. In this spectrum, three peaks are visible all of which can be attributed to beta-Ta peaks. There is a peak at 35.6° which is fitting for beta-Ta (321) and there are two more peaks at 59.3° and a very pronounced peak at 62.3° which are at the positions of beta-Ta (313) and (403), respectively. Practically no peaks are present at the positions where alpha-Ta peaks are expected. In the spectra of the tantalum sputtered at 20°C (b), black line, a peak is visible at the position of alpha-Ta (100). This peak is not very pronounced and it also appears as if the peak is flattened or there are two peaks merged.

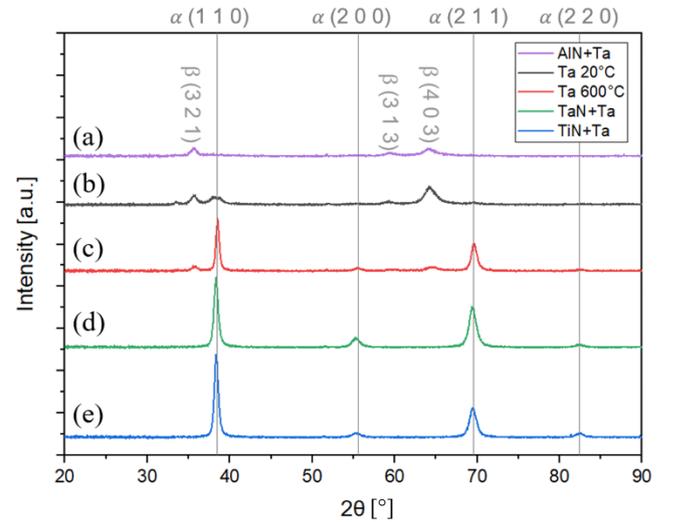

*Figure 2: X-ray diffraction spectra measured with GI-XRD with a grazing incidence angle of 0.5° for the Ta (a) on AlN seed layer, (b) on unheated Si, (c) on 600°C heated Si, (d) on TaN and (e) on TiN seed layer.*

There is also a very small peak visible where alpha-Ta (211) is supposed to be while a somewhat more pronounced beta-Ta (403) peak suggests a mixed-phase thin film. The red line in the middle (c) is the spectra of Ta sputtered on a silicon substrate heated to 600°C. The alpha-Ta peaks are here sharp and more pronounced, especially (110) and (211), but there are also small remaining beta-Ta peaks visible. There are also smaller peaks for alpha-Ta (200) and (220) present. The spectrum for tantalum on tantalum nitride (d) is shown below in green and for tantalum on titanium nitride in blue (e). Both spectra have four peaks which all lie at positions where the alpha-Ta is supposed to appear. This indicates that nucleation of the alpha-phase of Ta worked best with the TaN and the TiN seed layer, and that there is no visible beta-Ta content in these thin films left. The spectra appear very similar with only differences in the peaks for alpha-Ta (110) and (220) which are both more pronounced and sharper in the spectrum for the Ta film with the TiN seed layer below.

*B. Electrical characterization*

The results of the resistivity measurements for all investigated Ta thin-films are visualized in Fig. 3 (a). The green band at the bottom indicates the range of resistivity values anticipated for alpha-Ta (20-80 µΩ.cm).

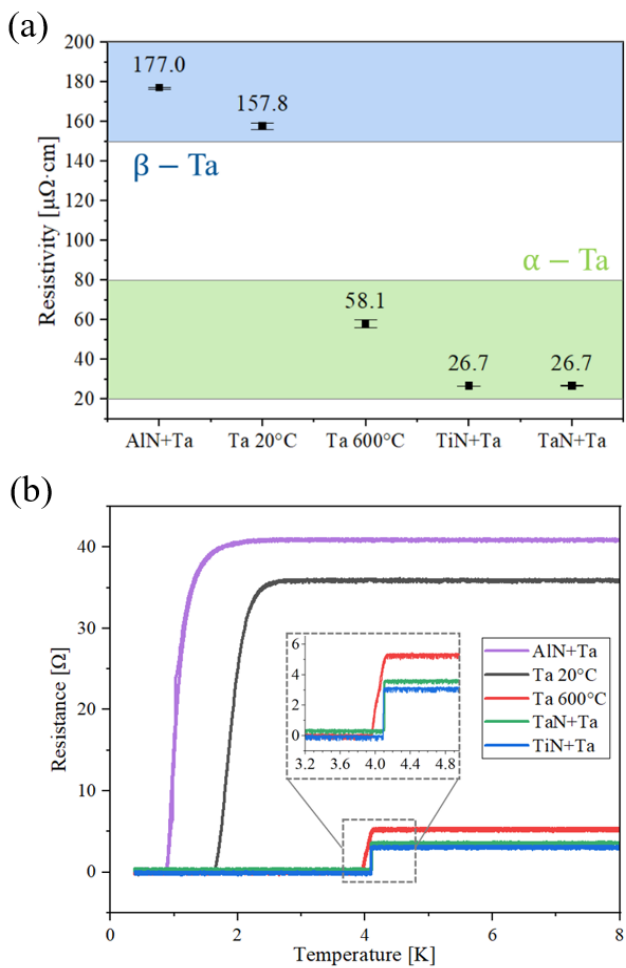

*Figure 3: DC electrical characterization of sputtered tantalum thin films. (a) Resistivity values of all investigated thin films; the blue region indicates beta-Ta, the green one alpha-Ta values, as derived from literature. (b) Critical temperature measurements: Resistance as function of temperature for all thin film samples, showing characteristic superconducting transitions at different temperatures and with different slopes. The inset shows the transition for the three (predominantly) alpha-Ta thin films in detail, where sharp transitions around 4.1 K can be seen.*

Likewise, the blue band indicates the range for beta-Ta (150-200 µΩ.cm) [27]. The tantalum that was sputtered onto the aluminum nitride seed layer shows high resistivity of 177.0 ± 0.4 µΩ.cm, which is in the reported range for beta-Ta, The value for the tantalum thin film that was sputtered without heating directly onto the silicon substrate has a resistivity of 157.8 ± 1.7 µΩ.cm. This value lies in the range of reported beta-Ta values, too, which is also in line with the XRD-spectrum of that thin film, showing a mix of phases dominated by beta-Ta contributions. The tantalum thin film sputtered on the 600°C heated substrate has a far lower resistivity of 58.1 ± 2.0 µΩ.cm, which lies higher up in the reported alpha-Ta value region. Finally, the value for the tantalum thin film on the titanium nitride seed layer is with 26.7 ± 0.2 µΩ.cm very similar to the one for the Ta on the tantalum nitride seed layer (26.7 ± 0.1 µΩ.cm). Both show resistivity values which match the literature reference for alpha-Ta well.

The same Hall-bar was then used to determine if and at which critical temperature $T_c$ the thin films would turn superconducting. The results from these measurements are shown in Figure 3 (b). All of the investigated tantalum films turned superconducting. The curve with the highest residual resistance before the transition is the one for the tantalum thin film that was deposited onto aluminum nitride. The superconducting transition does not occur suddenly at a defined temperature but is rather a transition that starts at around 2 K and has a finite slope until the resistance drops to zero just below 1 K. The curve for the tantalum sputtered at 20°C directly on silicon looks similar with a transition that has also a finite slope at the beginning. In this case, the residual resistance is lower, and the transition, which begins around 2.3 K, is not as steep as the one for Ta on AlN. The transition terminates at around 1.8 K, where the resistance turns zero. The remaining three thin films display a different behaviour. The tantalum sputtered at 600°C on silicon has a sharper transition that has still an inclination and occurs at a higher temperature of 4.16 K, which is the temperature where the thin film shows zero resistance, see inset in Figure 3 (b). The residual resistance above $T_c$ is much lower than that of the previously discussed samples. The tantalum thin films on the titanium nitride and tantalum nitride seed layers display sharp transitions with critical temperatures of 4.08 K and 4.11 K, respectively. These three films, which have already shown high or even pure alpha-Ta content both, in the XRD and resistivity measurements, hence feature also $T_c$ values that are close to the literature value for thin film alpha-Ta, which is around 4.4 K [28].

To determine the RRR values for each film, the value for the room-temperature (293 K) resistance was divided by the residual resistance just above the superconducting transition. The tantalum sputtered at 20°C and the one deposited on AlN both have an RRR of ~one, as their resistance was practically not decreasing at low temperatures. The tantalum thin film sputtered at 600°C has a room temperature resistance that is almost three times as high as the value above the transition, resulting in an RRR of 2.88. The tantalum sputtered on the TiN seed layer has an RRR of 2.19 while the value for TaN on Ta is 2.11. The highest measured RRR for Ta sputtered at 600°C indicates that likely the least density of impurities or defects is present in this film, due to the film growth at high temperatures, closely followed by the Ta films grown on TiN and TaN seed layers.

## C. Resonator measurements

One of each of the five different tantalum thin films was patterned into a CPW resonator as outlined in the previous section. A comparison of internal quality factors $Q_i$ of all of the different tantalum thin films is shown in Figure 4. The tantalum sputtered on the aluminum nitride seed layer has the lowest $Q_i$ value over all, which is also not power dependent. In the single-photon regime, the film has a $Q_i$ value of $1.7 \pm 0.5 \times 10^4$ and at high power (-90dBm) $1.7 \pm 0.5 \times 10^4$. Slightly above lies the tantalum that was sputtered without heating at room temperature on the silicon substrate. Here the $Q_i$ values are slightly higher and are also not power dependent, reaching $6.2 \pm 1.6 \times 10^4$ in the single-photon regime and $6.7 \pm 1.7 \times 10^4$ at high power. Next is the tantalum that was deposited on the tantalum nitride seed layer. The quality factors are power dependent with $1.4 \pm 0.3 \times 10^5$ in the single-photon regime and $4.8 \pm 1.4 \times 10^5$ at high power. In terms of electrical properties and diffraction spectra the Ta on the TiN seed layer was very similar to the Ta on TaN, see above. However, in the resonator measurements the Ta on TiN performs noticeably better than Ta on TaN. The tantalum on the titanium nitride reaches $Q_i$ values of $5.0 \pm 0.4 \times 10^5$ in the single-photon regime and $3.1 \pm 1.0 \times 10^6$ at high power. Both, the Ta on TaN and Ta on TiN films show characteristic power dependent progression of $Q_i$ being higher at high powers and lower at low power, likely due to the increase in saturation of TLS losses with higher powers [26]. The tantalum thin film that performs best in terms of internal quality factor is the one sputtered on the 600°C heated substrate. Here, the $Q_i$ values reach one million in the single photon regime with $1.0 \pm 0.2 \times 10^6$ and go up to $6.2 \pm 1.7 \times 10^6$ at high power. This tantalum film already revealed in the electrical and crystallographic characterization that it contains a high fraction of alpha-Ta, but that also some remaining beta-Ta is present. For example, there were peaks in the diffraction pattern that can only be attributed to beta-Ta. Nevertheless, it appears that this film performs best amongst all investigated tantalum thin films with respect to RF characteristics. A reason for this could be that the high temperature during the sputtering process causes less impurities and defects in the material and at the interface compared to sputtering at room temperature. From an electrical and crystallographic point of view, the Ta on seed layers TaN and TiN look purest in terms of alpha-Ta content, but they have one additional interface more than the Ta that is sputtered directly onto the silicon. Even though the chips were not taken out of the vacuum and were sputtered in-situ with the seed layer and the Ta on top, this additional interface could host additional sources for loss channels. The difference in $Q_i$ for Ta on TiN and Ta on TaN also shows that the losses that stem from the seed layer may not only be attributed to the interface between the seed layer and the tantalum but also from the nature of the seed layer itself. The TiN seed layer recipe was optimized to produce stoichiometric TiN, while for the TaN the same nitrogen to argon setting was used. Hence, the TaN seed layer may host more TLS or other defects causing losses than the stoichiometric TiN seed layer. This may in turn have resulted in a relatively lower performance of the Ta resonator on top.

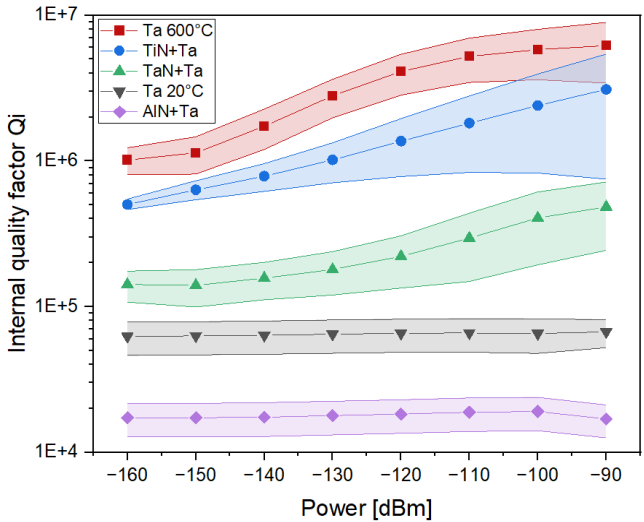

*Figure 4: Internal quality factors $Q_i$ for different tantalum thin films, as function of RF power. The tantalum thin film sputtered on the 600°C heated Si substrate performs best with $Q_i$ reaching one million at low powers. Symbols represent average data obtained from 8 resonators on one chip each, connected by full lines as guide to the eye. The surrounding shaded areas represent the standard deviation. Possible contributions from variations in $Q_{ext}$ were not taken into account for this analysis.*

## IV. CONCLUSION

Tantalum thin films were sputter-deposited on silicon substrates at different temperature or employing different seed layers. Crystallographic characterization shows that sputtering with the before-mentioned parameters at room-temperature leads to a mixture of alpha- and beta-phase in the tantalum films, which is in accordance with previous results. Increasing the temperature, while keeping all other parameters the same, yields tantalum thin films that comprise a high amount of alpha-phase. As alternative strategy, few nanometers thick seed layers were deposited onto the silicon substrate before the actual tantalum deposition. In the case of an aluminum nitride seed layer, the resulting tantalum thin film on top shows a high content of beta-tantalum. Using titanium nitride or tantalum nitride results in uniform, apparently all-alpha-tantalum films. Interestingly, the thin films containing mostly the beta-tantalum phase have low roughness (<1 nm RMS) while the alpha-tantalum films show higher values of 1.5-2 nm, approximately. Electrical measurements also indicate that the Ta on AlN is purely beta-Ta, the Ta sputtered on the unheated Si is also mostly beta-Ta while the Ta on the heated substrate is mostly alpha-Ta. The Ta on the TiN and TaN seed layers have typical values for both, resistivity and $T_c$, indicative of pure alpha-Ta films. All of the sputtered thin films turn superconducting, with critical temperatures for the alpha-Ta thin films around 4.1K and for the beta-Ta thin films at lower temperatures (~0.7 K - 2 K). CPW resonators measurements showed that the (almost) pure alpha-Ta thin films perform significantly better than the beta-Ta-dominated films in terms of internal quality factors $Q_i$ – in good overall agreement to the finding from structural and DC electrical characterization. In the case of Ta thin films on AlN, it should be noted that the low quality factor could also be partially attributed to the piezoelectricity of AlN [29]. The best performing thin-film, namely tantalum sputtered on 600°C heated silicon, reached an average value of about $Q_i = 1 \times 10^6$ in the single-photon regime. Even though this film comprised a somewhat less uniform alpha-phase, according to the diffraction spectrum and the electrical characterization, it even surpassed the Ta on TiN and on TaN. The reason for this could be that the additional interfaces between the Ta thin films and the seed layers may likely host more TLS, which would affect the performance of the otherwise highly uniform alpha-Ta films that grow on top of these seed layers. Further analysis based on different seed layer materials and thicknesses is currently in progress.

ACKNOWLEDGMENT

We acknowledge funding through the Munich Quantum Valley (MQV) project by the Free State of Bavaria, Germany. We would like to thank our colleagues from the TUM Catalysis Research Center (CRC) for giving us access to XRD-measurement equipment, the TUM Center for Nanotechnology and Nanomaterials (ZNN) for shared use of the clean room facilities, the technical staff working at TUM ZEITlab, and the colleagues at the Walther-Meissner-Institute for the initial resonator layout and copper box designs.